\documentclass[aps,twocolumn,prb,showpacs,floatfix]{revtex4}

\usepackage{epsfig,amsmath,amssymb}
\usepackage{graphics} 
\usepackage{subfigure}
\usepackage{color}
\usepackage{multirow}

\begin{document}

\title
{
Complete phase diagram of the spin-$\frac{1}{2}$ $J_{1}$--$J_{2}$--$J_{3}$ model 
(with $J_{3}=J_{2}$) on the honeycomb lattice
}

\author
{P.~H.~Y.~Li and R.~F.~Bishop}
\affiliation
{School of Physics and Astronomy, The University of Manchester, Manchester, M13 9PL, UK}


\begin{abstract}

  We use the coupled cluster method to investigate the ground-state (GS)
  properties of the frustrated spin-$\frac{1}{2}$ $J_{1}$--$J_{2}$--$J_{3}$ model 
  on the honeycomb lattice, with nearest-neighbor exchange coupling
  $J_1$ plus next-nearest-neighbor ($J_2$) and next-next-nearest-neighbor ($J_3$)
  exchanges of equal strength.  In particular we find  
  a direct first-order phase transition between the N\'{e}el-ordered 
  antiferromagnetic phase and the ferromagnetic phase at a value
  $J_{2}/J_{1} = -1.17 \pm 0.01$ when $J_{1}>0$, compared to the
  corresponding classical value of $-1$.  We find no evidence for any 
  intermediate phase.  From this and our previous CCM studies of the
  model we present its full zero-temperature GS phase diagram.

\end{abstract}

\maketitle

\section{Introduction}
Frustrated quantum spin systems on regular two-dimensional (2D) 
lattices have been the subject of intense interest in recent 
years.\cite{2D_magnetism_1,2D_magnetism_2,2D_magnetism_3}  They
exhibit a wide variety of different types of ordering and phases,
even at zero temperature ($T=0$).  Examples include various quasiclassical
antiferromagnetic (AFM) phases (e.g., with N\'eel or columnar
striped ordering), phases with quantum spiral ordering, valence-bond
crystalline phases with nematic ordering, and spin-liquid phases.
Of particular interest are the ($T=0$) quantum phase transitions
that can occur as the coupling constants in the Hamiltonian
are varied, so that the degree of frustration between bonds competing
for various types of order is changed.  The resulting interplay 
between magnetic frustration and quantum fluctuations has
been seen to be a very effective means to create (and destroy) new
types of order not present in the classical counterparts of
the models.  The successful syntheses of ever more quasi-2D
magnetic materials, and the experimental investigation of their 
properties, has also served to intensify their theoretical study.
The very recent prospects of being able to realise spin-lattice models
with ultracold atoms trapped in optical lattices\cite{Struck:2011} is
likely to make even more data available about the quantum phase transitions
in the models as the exciting possibility opens up in such trapped-atom
experiments to tune the strengths of the competing magnetic bonds,
and hence to drive the system from one phase to another.

Since quantum fluctuations tend to be largest for the smallest values of the 
spin quantum number $s$, for lower dimensionality $D$ of the lattice, and for 
the smallest coordination number $z$ of the lattice, spin-$\frac{1}{2}$ models on the
(hexagonal or) honeycomb lattice play a special role for $D=2$, since 
the honeycomb lattice has the lowest $z$ ($=3$) of
all regular 2D lattices.  Thus, for example, one of the few exactly solvable models on
the honeycomb lattice, namely the Kitaev model,\cite{kitaev} has been 
shown to sustain a spin-liquid phase.  Clearly, the honeycomb lattice
is also relevant to the study of graphene, for which much of the
physics may be describable in terms of Hubbard-like models on
this lattice.\cite{graphene}
Evidence has also been found from quantum Monte Carlo (QMC) 
studies\cite{meng} that quantum fluctuations are sufficiently 
strong to establish an insulating spin-liquid phase between the nonmagnetic 
metallic phase and the antiferromagnetic (AFM) Mott insulator phase, when the 
Coulomb repulsion parameter $U$ becomes moderately strong.  For large values of 
$U$ the latter phase corresponds to the pure Heisenberg antiferromagnet (HAFM) 
on the bipartite honeycomb lattice, 
whose GS phase exhibits N\'{e}el LRO.  However, higher-order terms in the $t/U$
expansion of the Hubbard model may lead to frustrating exchange couplings 
in the corresponding spin-lattice limiting model, in which the HAFM with 
nearest-neighbor (NN) exchange
couplings is the leading term in the large-$U$ expansion.  
Frustration is easily incorporated via competing next-nearest-neighbor (NNN) and 
maybe also next-next-nearest-neighbor (NNNN) bonds.  
Recent calculations of the low-dimensional material $\beta$-Cu$_{2}$V$_{2}$O$_{7}$
also show that its magnetic properties can be described in terms of
a spin-$\frac{1}{2}$ model on a distorted honeycomb lattice.\cite{Tsirlin:2010} 

\begin{figure}[t]
\includegraphics[width=4cm]{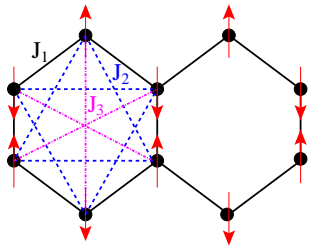}
\caption{(Color online) The N\'{e}el state and the bonds of the $J_{1}$-$J_{2}$-$J_{3}$ honeycomb model;
  the arrows represent spins located on lattice sites \textbullet.}
\label{model}
\end{figure}

For all these and other reasons, frustrated spin-$\frac{1}{2}$ Heisenberg models on the
honeycomb lattice, including couplings $J_1$, $J_2$, and $J_3$ up to third 
nearest neighbors, have been extensively studied using a variety of theoretical
tools.\cite{Rastelli:1979,Fouet:2001,Mulder:2010,Cabra:2011,Ganesh:2011,Clark:2011,DJJF:2011_honeycomb,Reuther:2011,
Lauchli:2011,Oitmaa:2011,Mosadeq:2011,PHYLi:2012_honeycomb_J1neg,PHYLi:2012_honeyJ1-J2}
The Hamiltonian for the system is
\begin{equation}
H = J_{1}\sum_{\langle i,j \rangle} \mathbf{s}_{i}\cdot\mathbf{s}_{j} + 
J_{2}\sum_{\langle\langle i,k \rangle\rangle} \mathbf{s}_{i}\cdot\mathbf{s}_{k} + 
J_{3}\sum_{\langle\langle\langle i,l \rangle\rangle\rangle} \mathbf{s}_{i}\cdot\mathbf{s}_{l}\,,   
\label{H}
\end{equation}
where $i$ runs over all lattice sites on the lattice, and $j$ runs
over all NN sites, $k$ over all NNN sites, and $l$ over all NNNN
sites to $i$, respectively, counting each bond once and once
only. Each site $i$ of the lattice carries a particle with spin operator 
${\bf s}_{i}$ and spin quantum number $s = \frac{1}{2}$.
The lattice and the exchange bonds are illustrated in Fig.~\ref{model}.

The solution of the classical version of the model (i.e., when $s \to \infty$) is itself 
rich.\cite{Rastelli:1979,Fouet:2001}  For the AFM version of the model (i.e.,
when $J_{1}>0$) there are six different ground-state (GS) phases, 
comprising three different collinear AFM phases, the ferromagnetic (FM) state, and two different 
helical phases (and see, e.g., Fig.~2 of Ref.~\onlinecite{Fouet:2001}).  
The AFM phases are the N\'{e}el phase (N) shown in Fig.~\ref{model}, and the 
so-called striped (S) and anti-N\'{e}el (aN) phases. 
The S, aN, and N states have, respectively, 1, 2, and all 3 NN spins
to a given spin antiparallel to it.  Equivalently, if we consider the sites of the honeycomb lattice
to form a set of parallel sawtooth (or zigzag) chains (in any one of the three equivalent
directions), the S state comprises alternating FM chains, while the aN state comprises AFM chains 
in which NN spins on adjacent chains are parallel.  There are actually infinite
manifolds of non-coplanar states degenerate in energy with each of the S and aN states at $T=0$, 
but both thermal and quantum fluctuations select the collinear configurations.\cite{Fouet:2001} 
At the classical level there is an exact symmetry between the GS phase diagrams of the
AFM ($J_{1}>0$) and FM ($J_{1}<0$) models, whereby one maps into the other under the
interchanges $J_{1} \rightleftharpoons -J_{1}$, $J_{3} \rightleftharpoons -J_{3}$, and
${\bf s}_{i} \rightleftharpoons -{\bf s}_{i}$, for sites $i$ belonging to one of the two
equivalent triangular sublattices of the honeycomb lattice (c.f., Figs.~2 and 3 of
Ref.~\onlinecite{Fouet:2001}).  When all three
bonds are AFM in nature, the three possible GS phases (viz., the N, S, and 
one of the helical phases) meet at a tricritical point at
$J_{3}=J_{2}=J_{1}/2$.  

The line $J_{3}=J_{2}$ ($\equiv \alpha J_{1}$) is thus of special
interest, and we henceforth restrict ourselves to this situation where the NNN
and NNNN bonds have equal strength for the remainder of the paper.
There are then 4 GS classical phases.
For the AFM case (with $J_{1}>0$) we have: (a) the AFM S state when
$\alpha > \frac{1}{2}$; (b) the AFM N state when $-1 < \alpha < \frac{1}{2}$;
and (c) the FM state when $\alpha < -1$.  For the FM case (with $J_{1} < 0$)
we have: (a) the FM state when $\alpha > -\frac{1}{10}$; (b) a spiral state
when $-\frac{1}{5} < \alpha < -\frac{1}{10}$; and (c) the AFM S state
when $\alpha < -\frac{1}{5}$.  The N state has classical first-order
phase transitions to both the S state and the FM state, while the spiral 
state has continuous transitions to both the S state and the FM state.

\begin{figure}[!t]
\begin{center}\includegraphics[width=6cm,angle=270]{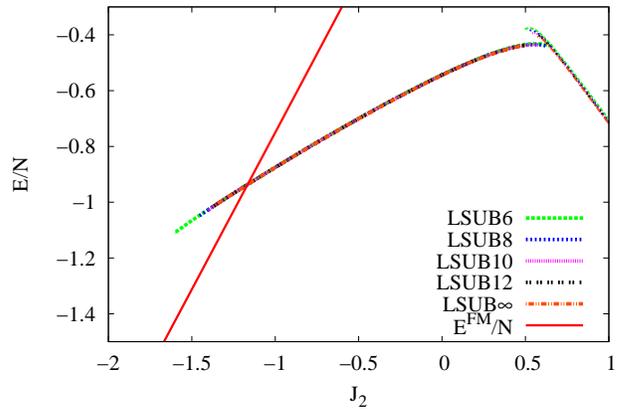}
\caption{(Color online) CCM LSUB$m$ results for the GS energy per spin, $E/N$, 
($J_{1}\equiv 1$ and $J_{3}=J_{2}$) based on the N\'{e}el state (left thick curves)
and striped state (right thin curves) as model states. We show calculated results
for $m=\{6,8,10,12\}$ and the extrapolated LSUB$\infty$ result (see text). The exact FM result,
$E^{\rm FM}/N=3(1+3J_{2})/8$, is also shown.}
\label{E}
\end{center}
\end{figure} 

In two previous papers\cite{DJJF:2011_honeycomb,PHYLi:2012_honeycomb_J1neg} 
we have applied the coupled cluster method (CCM) to the $s=\frac{1}{2}$ version 
of this model (with $J_{3}=J_{2}$).  For the AFM case (with $J_{1}>0$) 
we found\cite{DJJF:2011_honeycomb}
that the direct classical first-order phase transition between the two (N and S) AFM states
at $\alpha_{\rm cl}^{\rm N-S}=0.5$ is changed for the quantum $s=\frac{1}{2}$ model 
into two separate transitions, so that a N\'{e}el-ordered (N) phase
exists for $\alpha < \alpha_{c_1} \approx 0.47$ and an AFM stripe-ordered (S)
phase exists for $\alpha > \alpha_{c_2} \approx 0.60$.  In between, for
$\alpha_{c_1} < \alpha < \alpha_{c_2}$, we found a paramagnetic GS phase
with plaquette valence-bond crystalline (PVBC) ordering that has no 
classical counterpart.  We further found that the quantum critical point (QCP) at
$\alpha_{c_2}$ appears to be first order, while that at $\alpha_{c_1}$ is continuous.
Since the N and PVBC phases break different symmetries we argued that our 
results favor the deconfinement scenario\cite{Senthil:2004} for the latter transition.
For the FM case (with $J_{1}<0$) we found\cite{PHYLi:2012_honeycomb_J1neg} that the two classical
transitions from the spiral phase to the S phase at $\alpha_{\rm cl}^{\rm sp-S}=-0.2$ and
from the spiral phase to the FM phase at $\alpha_{\rm cl}^{\rm sp-FM}=-0.1$ are 
changed for the quantum $s=\frac{1}{2}$ model into one of two scenarios, namely, 
either a direct first-order transition
between the the AFM S state and the FM state at $\alpha_{c_3} \approx -0.11$, or
there exists an intervening phase between them in the very narrow range
$-0.12 \lesssim \alpha \lesssim -0.10$.

In order to complete the phase diagram of the spin-$\frac{1}{2}$ model there remains
to investigate the quantum analog of the classical first-order transition at 
$\alpha_{\rm cl}^{\rm N-FM}=-1$ between the AFM N state and the FM state for the
AFM case (with $J_{1}>0$), and that is the purpose of this paper.  We
shall again use the CCM to do so.  The two
states for the classical model have respective energies per spin given by
$E_{\rm cl}^{\rm N}/N = \frac{3}{2}s^{2}(-J_{1}+J_{2})$ and
$E_{\rm cl}^{\rm FM}/N = \frac{3}{2}s^{2}(J_{1}+3J_{2})$.  Hence at the phase transition
point $\alpha_{\rm cl}^{\rm N-FM}=-1$ the GS energy per spin is 
$E_{\rm cl}/N = - \frac{3}{4}$ if we take $s=\frac{1}{2}$ and $J_{1}=+1$ to
set the energy scale.  In all that follows we take $J_{1}\equiv 1$.
\begin{figure}[!t]
\includegraphics[width=6cm,angle=270]{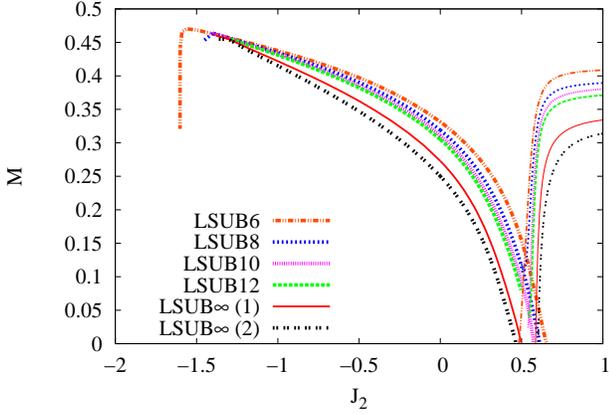}
\hfil
\caption{(Color online) CCM LSUB$m$ results for the GS order parameter, $M$,
($J_{1}\equiv 1$ and $J_{3}=J_{2}$) based on the N\'{e}el state (left thick curves)
and striped state (right thin curves) as model states. We show calculated results
for $m=\{6,8,10,12\}$ and the extrapolated LSUB$\infty(1)$ 
and LSUB$\infty(2)$ results (see text).}
\label{M}
\end{figure}

The CCM (see, e.g., Refs.~\onlinecite{Bi:1991,Zeng:1998,Fa:2004} and
references cited therein) is one of the most powerful and most
versatile modern techniques in quantum many-body theory.  It has been
applied to many quantum magnets with huge success (see
Refs.~\onlinecite{DJJF:2011_honeycomb,PHYLi:2012_honeycomb_J1neg,PHYLi:2012_honeyJ1-J2,Fa:2004,rachid05,Bi:2008_JPCM,square_triangle}
and references cited therein).  The interested reader  
can find details of the CCM in the references cited, and we do not elaborate here.
We note only that it is a size-extensive method that provides
results from the outset in the infinite-lattice limit ($N \to \infty$).  The method requires
us to provide a model (or reference) state, with respect to which the
quantum correlations are expressed.  Here we simply use the N state
shown in Fig.~\ref{model}, although for comparison purposes we also display
below results obtained previously\cite{DJJF:2011_honeycomb} based on the S state.  
As before, we use the well-tested localized  
lattice-animal-based subsystem (LSUB$m$) truncation scheme in which all
multispin correlations are retained in the CCM correlation operators over 
all distinct locales on the lattice defined by $m$ or fewer contiguous sites.
The method of solving for higher orders of LSUB$m$ approximations is discussed 
in detail in Ref.~\onlinecite{Zeng:1998}.
\begin{figure}[!t]
\begin{center}
\vskip-0.8cm
\includegraphics[width=9cm,angle=90]{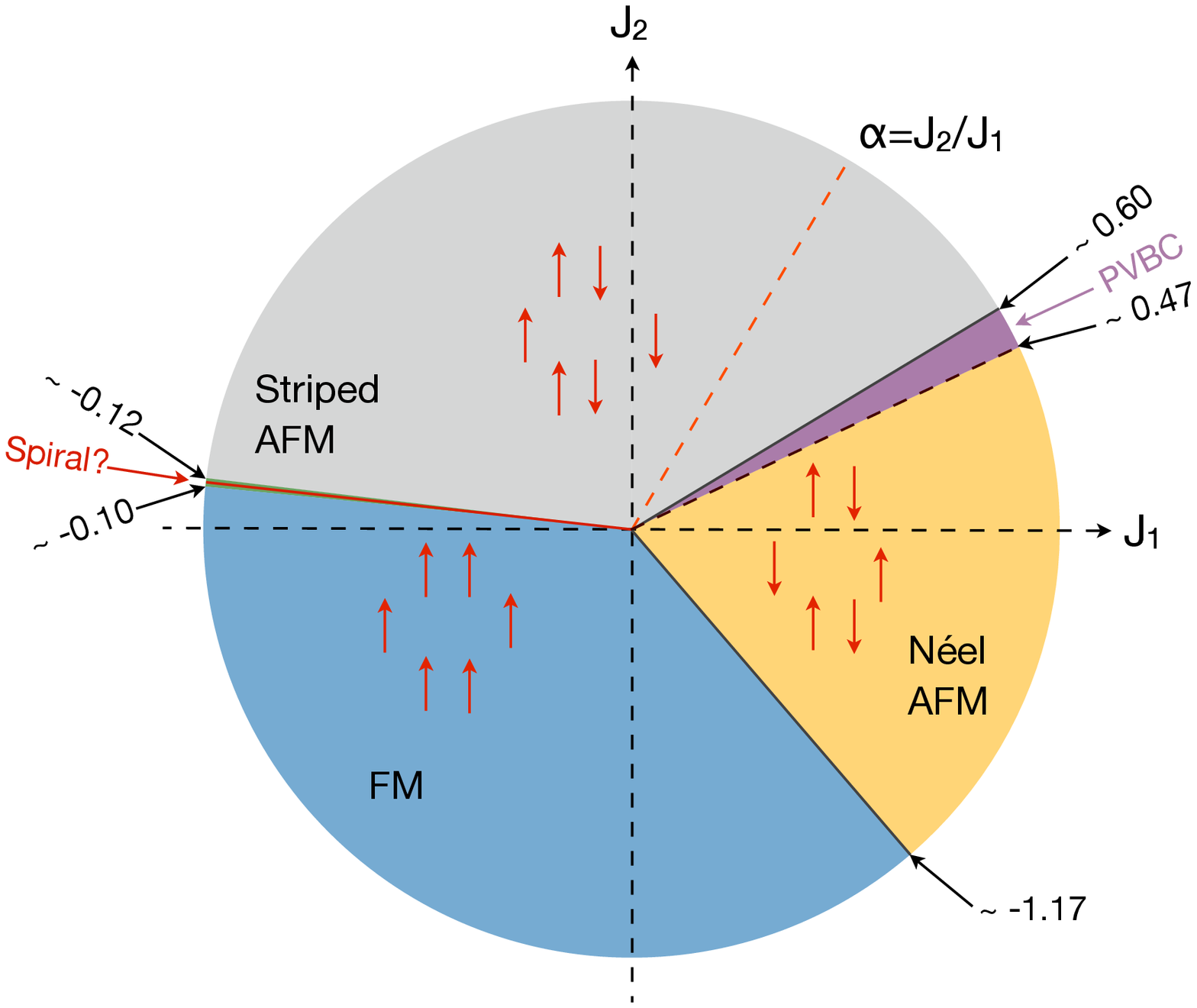}
\vskip-1.5cm
\caption{(Color online) The phase diagram of the spin-$\frac{1}{2}$ 
$J_{1}$--$J_{2}$--$J_{3}$ honeycomb model 
in the $J_1$-$J_2$ plane, for the case $J_{3}=J_{2}$.  
The continuous transition between the N\'eel  and PVBC phases at
$J_{2}/J_{1} \equiv \alpha = \alpha_{c_1} \approx 0.47$ is shown by a broken line, while the 
first-order transition between the PVBC and striped phases
at $\alpha_{c_2} \approx 0.60$ is shown by a solid line.  The transition
between the striped and FM phases is either a first-order 
one at $\alpha_{c_3} \approx -0.11$, shown by a solid line, or occurs via an intermediate
phase, probably with noncollinear spiral order, which exists 
in the region $-0.12 \lesssim \alpha \lesssim -0.10$.  The first-order transition
between the FM and the N\'eel phases at $\alpha_{c_4} \approx -1.17$
is shown by a solid line.}
\label{phase-diagram}
\end{center}
\end{figure}

The number of independent fundamental clusters 
increases rapidly with the LSUB$m$ truncation
index $m$.  Hence, it is essential to employ parallel processing
techniques and supercomputing resources for larger values 
of $m$.\cite{ccm}  To obtain results in the (exact) $m \to \infty$
limit, we need to extrapolate the raw LSUB$m$ data.  Since the hexagon is
a fundamental element of the honeycomb lattice we use LSUB$m$ data only with 
$m \geq 6$.  For the GS energy per spin we employ the usual and well-tested 
scheme, $E(m)/N = a_{0}+a_{1}m^{-2}+a_{2}m^{-4}$.
For the magnetic order parameter (or average onsite magnetization), $M$, different schemes have been
used for different situations.  For models with no or only little frustration a well-tested
scheme is $M(m) = b_{0}+b_{1}m^{-1}+b_{2}m^{-2}$, whereas a more appropriate scheme for
highly frustrated models, especially those showing a GS quantum phase transition, is
$M(m) = c_{0}+c_{1}m^{-1/2}+c_{2}m^{-3/2}$.  We henceforth refer to these latter two schemes
for $M$ as LSUB$\infty(1)$ and LSUB$\infty(2)$, respectively. 
All of the the extrapolations shown below are based on LSUB$m$ results
with $m=\{6,8,10,12\}$.  

\section{Results}
\label{results}
In Fig.~\ref{E}
we show our CCM results for the GS energy per spin, $E/N$.  They are
evidently very well converged for all values of $J_2$ shown.  There is
a clear energy crossing of the FM and (extrapolated) N energy curves at a value
$\alpha_{c_4} \approx -1.17$, with $E/N \approx -0.941$, which is direct
evidence of a first-order phase transition, just as in the classical case where it occurs
at $\alpha_{\rm cl}^{\rm N-FM}=-1$ with $E_{\rm cl}/N=-0.75$.

We note that the individual LSUB$m$ energy curves based on the N\'{e}el model
state terminate at some lower critical value, $\alpha_{t}^{{\rm LSUB}m}$ which itself depends on the
index $m$.  These termination points, below which no real solutions to 
the coupled CCM equations exist, are themselves a reflection of the actual
QCP at $\alpha_{c_4}$.  For example, $\alpha_{t}^{{\rm LSUB}12} \approx -1.38$. 
In Fig.~\ref{M} we show the corresponding results for the GS magnetic order
parameter, $M$.  We observe that the behavior of $M$ on the N\'{e}el side near the QCP at 
$\alpha_{c_4}$ is quite smooth, with the only indication of the
transition to the FM state being the downturn very near the
$\alpha_{t}^{{{\rm LSUB}}m}$ termination points.  This is in sharp
contrast to the behavior at the other end near the QCP of the N state with the 
PVBC state at $\alpha_{c_1}$ where $M \to 0$.  Clearly the best estimate for 
$\alpha_{c_1}$ comes from the LSUB$\infty(2)$ extrapolation, whereas the best
estimate for $M$ at the pure honeycomb HAFM point (i.e., when $J_{3}=J_{2}=0$)
comes from the LSUB$\infty(1)$ extrapolation, which gives $M = 0.272 \pm 0.002$,
in excellent agreement with the value $M=0.2677 \pm 0.0006$ from a QMC simulation
of lattices up to size $N=2048$.\cite{Castro:2006}  Figure \ref{M} also
clearly shows the corresponding transition at $\alpha_{c_3}$ between the S state
and the PVBC state, where again $M \to 0$ on the striped phase side.

\section{Summary and Conclusions}
\label{conclusions}
In this and previous
papers,\cite{DJJF:2011_honeycomb,PHYLi:2012_honeycomb_J1neg} we have
studied the spin-$\frac{1}{2}$ $J_{1}$--$J_{2}$--$J_{3}$ Heisenberg
model, with $J_{2}=J_{3}$, on the honeycomb lattice, using the CCM.
In the present paper we have concentrated on completing the phase
diagram.  In particular we find that the classical direct first-order
phase transition for the AFM case (where $J_{1}>0$) between the AFM
N\'{e}el-ordered phase and the FM phase is preserved for the quantum
spin-$\frac{1}{2}$ model, but now occurs at a QCP, $\alpha_{c_4}
\approx -1.17 \pm 0.01$, compared to the classical value 
$\alpha_{\rm cl}^{\rm N-FM}=-1$.  Thus quantum fluctuations act to stabilize the
collinear AFM order at the expense of the FM order, to higher values
of frustration than in the classical case, as has also been observed
in the FM version of the spin-$\frac{1}{2}$ $J_{1}$--$J_{2}$ model on
the square lattice.\cite{Richter:2010}  We find no evidence that
quantum fluctuations permit an intervening state with no classical
counterpart, unlike the case of the transition between the two
(N\'{e}el-ordered and stripe-ordered) AFM states which occurs as a
direct first-order phase transition at $\alpha_{\rm cl}^{\rm N-S}=0.5$
for the classical model, but occurs in the quantum spin-$\frac{1}{2}$
model via the intermediate PVBC phase.

Our results from this and our previous CCM studies\cite{DJJF:2011_honeycomb,PHYLi:2012_honeycomb_J1neg} 
are summarised in the complete phase diagram for the model shown in Fig.~\ref{phase-diagram}.
 
We thank the University of Minnesota Supercomputing Institute for
Digital Simulation and Advanced Computation for the grant of
supercomputing facilities.  We also thank D.~J.~J.~Farnell and 
J.~Richter for fruitful discussions.

\end{document}